# Charge transport through redox active $[H_7P_8W_{48}O_{184}]^{33-}$ polyoxometalates self-assembled onto gold surfaces and gold nanodots.

K. Dalla Francesca,[1] S. Lenfant,[2] M. Laurans,[1] F. Volatron,[1] G. Izzet,[1] V. Humblot,[3] C. Methivier,[3] D. Guerin,[2] A. Proust,[1,]* D. Vuillaume[2,]*

1) Sorbonne Université, CNRS, Institut Parisien de Chimie Moléculaire, IPCM, 4 Place Jussieu, F-75005 Paris, France.
2) Institute for Electronics Microelectronics and Nanotechnology (IEMN), CNRS, Lille Univ., Av. Poincaré, F-59652 Villeneuve d'Ascq, France.
3) Sorbonne Université, CNRS, Laboratoire de Réactivité de Surface, LRS, 4 Place Jussieu, F-75005 Paris, France.

*Corresponding authors: anna.proust@upmc.fr, dominique.vuillaume@iemn.fr

**Abstract.** Polyoxometalates (POMs) are redox-active molecular oxides, which attract growing interest for their integration into nano-devices, such as high-density data storage non-volatile memories. In this work, we investigated the electrostatic deposition of the negatively charged $[H_7P_8W_{48}O_{184}]^{33-}$ POM onto positively charged 8-amino-1-octanethiol self-assembled monolayers (SAMs) preformed onto gold substrates or onto an array of gold nanodots. The ring-shaped $[H_7P_8W_{48}O_{184}]^{33-}$ POM was selected as an example of large POMs with high charge storage capacity. To avoid the formation of POM aggregates onto the substrates, which would introduce variability in the local electrical properties, special attention has to be paid to the preformed SAM seeding layer, which should itself be deprived of aggregates. Where necessary, rinsing steps were found to be crucial to eliminate these aggregates and to provide uniformly covered substrates for subsequent POM deposition and electrical characterizations. This especially holds for commercially available gold/glass substrates while these rinsing steps were not essential in the case of template stripped gold of very low roughness. Charge transport through the related molecular junctions and nanodot molecule junctions (NMJs) has been probed by conducting-AFM. We analyzed the current-voltage curves with different models: electron tunneling though the SAMs (Simmons model), transition voltage spectroscopy (TVS) method or molecular single energy level mediated transport (Landauer equation) and we discussed the energetics of the molecular junctions. We concluded to an energy level alignment of the alkyl spacer and POM lowest occupied molecular orbitals (LUMOs), probably due to dipolar effects.

# 1. Introduction.

Redox active molecules, which can be switched from one redox state to another, have been considered as promising for the development of molecular memory devices.[1, 2-4, 5, 6, 7] Often viewed as the missing link between conventional molecules and extended oxides, which are ubiquitous in electronics, polyoxometalates (POMs) should thus draw more attention. These molecular oxides obeying to the general formula $[X_xM_pO_y]^{n-}$ (X= P, Si …; M = Mo$^{VI}$, W$^{VI}$, V$^{V}$…) display an outstanding chemical versatility and remarkable redox properties.[8-13] There are still very few reports on single POM devices.[14-15] Charge transport through large POM-based molecular junctions prepared by layer-by-layer deposition or dip-coating on various electrodes has been studied[16] and a metal-insulator/POM-semiconductor capacitor cell has been reported.[17] As a major milestone, a flash-type memory cell comprising POMs drop-cast at the Si-nanowire channel has been recently described.[18] However, in all the above POM-based devices the molecular organization and the POM packing density are not controlled and this was found to confine the performances. Therefore special attention should be paid to the processing of POMs to get uniformly structured thus reproducible layered materials and to reduce device-to-device variability of the ultimate electrical properties which is a main issue in nanoelectronics.

Yet POM processing is still a sticking point. Direct deposition of POMs onto electrodes has been probed by microscopy imaging[19-22, 23-25] and largely exploited for electrocatalysis,[26-28] sensing purposes,[29] solar-energy conversion[30] and in composite materials for molecular batteries or supercapacitors.[31, 32-34, 35-37] Drop-casting[18, 38-39] and spin-coating of POM solutions are easy to carry out and have supplied, among others, interfacial layers for solar cells or light-emitting devices.[40-41, 42, 43-45] In those cases no special order in the POM arrangement was sought. As POMs are negatively charged species their immobilization onto electrodes has also commonly relied on their entrapment into positively charged polymers[46-48] or on the dip-coating exchange of their counter cations: by positively charged electrolyte through the Layer-by-Layer method to build photo- or electro-chromic materials,[49-50] photo-electrodes,[51] interfacial layers for solar cells[52-54] or multilayer films for electrocatalysis;[55-57] by amphiphilic cations to form Langmuir Blodgett films;[58-62] by positively charged groups directly decorating the electrode,[63] for application in water splitting,[64-68] molecular cluster batteries[69] or molecular electronics.[17, 70-71] Finally but to a lesser extent, covalent grafting of POMs onto electrodes, directly or in two steps, has been investigated by some of us and others.[72-76] This allowed us to describe the electron transfer kinetics from the electrode to a compact POM monolayer, assembled onto carbonaceous materials,[76-78] gold [79-80] or silicon wafers[81] and electron transport through silicon-POMs-metal junctions,[82] as a first step towards integration into nanoelectronic devices.

Although a covalent route provides a better control of the POMs/electrode interface, it is highly demanding in terms of synthetic efforts since it requires prior POM functionalization to obtain organic-inorganic POM hybrids with suitable remote anchor. For this reason, it has been restricted to the immobilization of POMs of the Anderson-, Lindqvist- and Keggin-types, since functionalization of larger POMs remains to be rationalized. On the other hand, an electrostatic route is easier to implement and allows drawing into the whole POM library, including larger POMs with higher charge storage capacity, for a rapid benchmarking. In this contribution, we thus investigated the immobilization of $[H_7P_8W_{48}O_{184}]^{33-}$ by dip-coating onto 8-amino-1-octanethiol self-assembled monolayers (SAMs) preformed onto Au-substrates or onto an array of Au-nanodots and we have characterized the charge transport through the resulting molecular junctions by C-AFM. Our choice of the $[H_7P_8W_{48}O_{184}]^{33-}$ POM is motivated by its high stability and also by its remarkable redox properties since, in solution and at an appropriate pH, its electrochemistry proceeds through up to three successive, 8-electron reduction waves.[83]

## 2. Deposition of $[H_7P_8W_{48}O_{184}]^{33-}$ onto 8-amino-1-octanethiol SAMs.

The immobilization of negatively charged POMs onto electrodes coated with SAMs terminated by positively charged groups is easy to implement but getting a uniformly structured layer not guaranteed. Deposition of $H_3[PW_{12}O_{40}]$ onto a Si/SiO$_2$ substrate covered by APTES by dip-coating resulted in the formation of nano-islands of POM aggregates, with a mean diameter of 17 nm and a height of 5-14 nm.[17] Hydrosilylation of undecylenic acid on hydrogenated silicon, followed by coupling to N,N-dimethylethylenediamine supplied amino-functionalized SAMs for subsequent deposition of $H_4[SiW_{12}O_{40}]$: mean surface coverage densities deduced from XPS experiments suggested the formation of monolayers but no details were given about the POM organization in the layers.[71] The electrochemically-driven deposition of $H_4[SiW_{12}O_{40}]$ on glassy-carbon electrode or HOPG modified by a 4-aminophenyl SAM gave close-packed ordered arrays according to STM images.[84] These previously published results point out that special attention should be paid to the experimental conditions for the solution processing of POMs. The POM concentration in the dipping solution and the incubation time are obviously key parameters but we have also particularly taken care of the quality of the seeding SAM. These parameters were probed by ellipsometry, PM-RAIRS and AFM for POMs deposited onto Au/Si (ellipsometry, PM-RAIRS) or Au/glass (AFM) respectively. Once the best conditions have been determined on commercially available substrates, POM layers have been prepared onto template-stripped Au surfaces ($^{TS}$Au)[85] and Au nanodots[86-87] for electrical characterization.

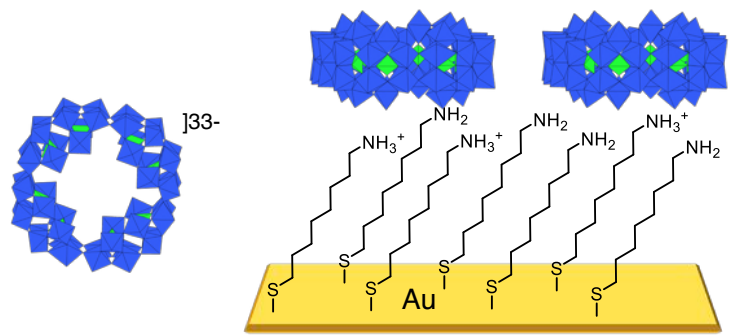

**Scheme 1.** *Polyhedral representation [$H_7P_8W_{48}O_{184}$]$^{33-}$(color code: $WO_6$ octahedra blue, $PO_4$ tetrahedra green) and schematization (not at scale) of its electrostatic deposition onto a 8-amino-1-octanethiol (AOT) SAM. The POM has a wheel-shaped structure with a diameter of ~2.0 nm and a thickness of ~1.0 nm.*

**Preparation of the seeding 8-amino-1-octanethiol SAM.**

The 8-amino-1-octanethiol (AOT) was chosen with an alkyl chain long enough to ensure a good organization of the SAM on gold but not too long to limit the decrease of the tunneling current through the layer.[88-91] The Au substrates were incubated in a $10^{-4}$ M solution of AOT hydrochloride in absolute ethanol for 24 h in the dark. Subsequent rinsing steps were found to be crucial to eliminate aggregates and to provide uniformly covered substrates for subsequent POM deposition. According to AFM monitoring (figure 1), our best conditions consisted in a five-step procedure: immersion in ethanol for 5 minutes, followed by ultra-sonication in ethanol for 5 minutes; immersion in 0.01 M Phosphate Buffer Saline (PBS pH = 7.4) for 3h, followed by ultra-sonication in distilled water for 5 minutes. The substrates were finally washed with ethanol and dried under nitrogen. The numerous ca. 2 nm sized aggregates observed on the AFM image (see Experimental Section) before PBS treatment (figure 1-a) disappeared after the 3h immersion in PBS (figure 1-b).

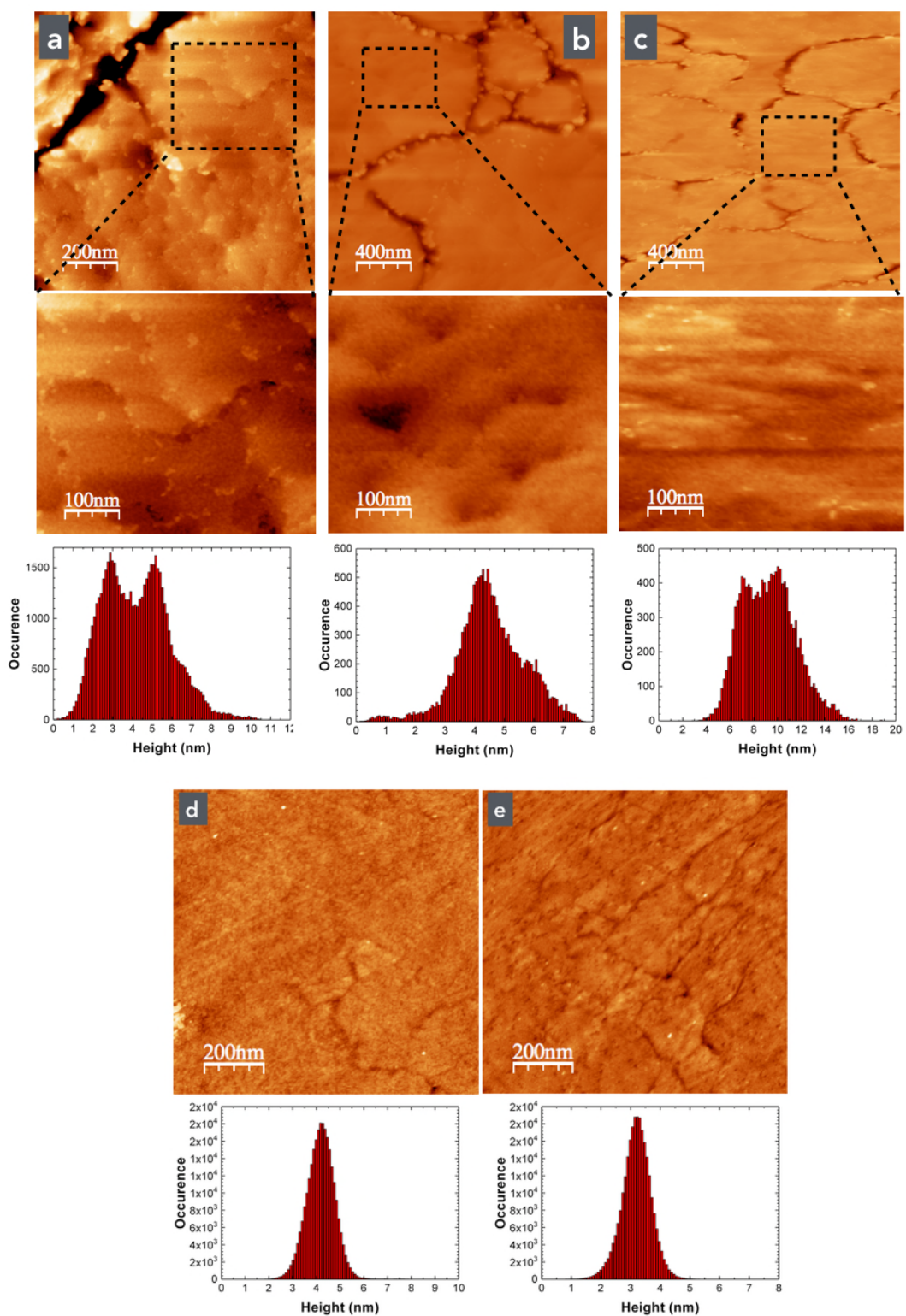

*Figure 1.* AFM images, zoom of the images (dashed line areas) and histograms of heights (from top to bottom) of the AOT SAMs on Au/glass substrates: **(a)** before and **(b)** after PBS immersion and **(c)** with the POM layer. AFM images (1x1 µm$^2$) and histograms of heights of the AOT SAMs on $^{TS}$Au substrates **(d)** without PBS rinsing and **(e)** with the POM layer. The histograms of heights clearly show the improvement of the surface topography for SAMs on the $^{TS}$Au electrodes, with a less *dispersed, single peak, distribution (d and e). The root mean square (rms) roughnesses, calculated from these height histograms, are **(b)** 1.27 nm, **(c)** 1.66 nm on the zoom images **(d )** 0.6 nm and **(e)** 0.5 nm.*

The AOT SAMs were also characterized by PM-RAIRS (see Experimental Section). The spectra (figure 2) displayed the characteristic bands corresponding to the vibrations of the alkyl chains (around 2920 and 2850 cm$^{-1}$ for the asymmetric and symmetric $\nu_{C-H}$ stretching modes and a broad band around 1400 cm$^{-1}$ for the $\delta_{C-H}$ cissor vibration) together with the bands assigned to the $\delta_{N-H}$ deformations at about 1650 and 1550 cm$^{-1}$ for the unprotonated and protonated amino groups, respectively (see Figure 2).

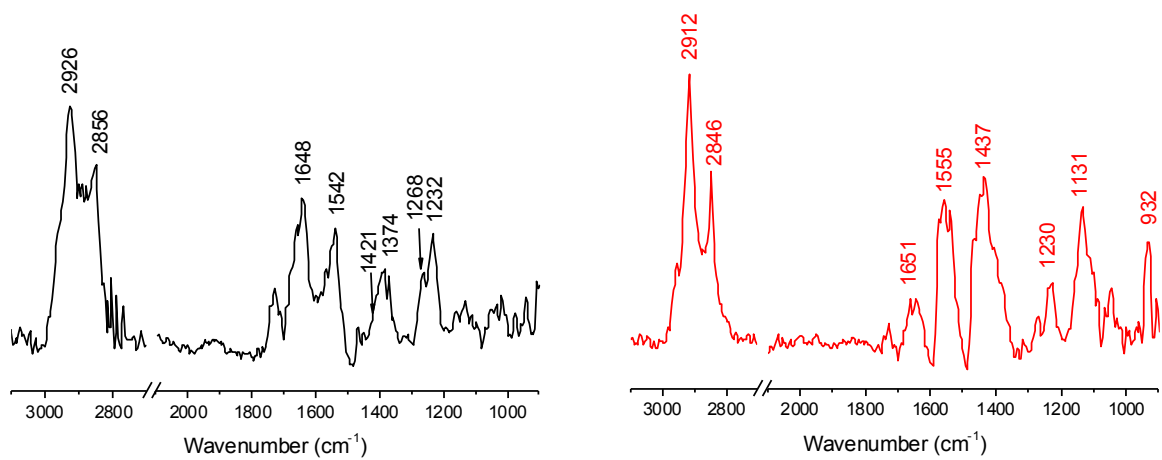

*Figure 2. PM-RAIRS spectra of the AOT functionalized Au substrate before (left) and after (right) POM deposition.*

**POM immobilization.**

The AOT functionalized Au substrates were then incubated in solutions of K$_{28}$Li$_5$[H$_7$P$_8$W$_{48}$O$_{184}$] in LiCl 2M and for various times. The solubility of K$_{28}$Li$_5$[H$_7$P$_8$W$_{48}$O$_{184}$] in water is poor but increased in the presence of LiCl. The deposition of the POM layer was monitored by ellipsometry, PM-RAIRS and AFM to select the optimal conditions, which were found to correspond to an incubation time of around 1 hour for the AOT functionalized gold substrate into a 10$^{-3}$ M POM solution, followed by washing with distilled water and ethanol and drying under nitrogen. Longer immersion times (until 24 h) led to exactly the same layer features and significantly more concentrated solutions were not possible to obtain because of the low solubility of the K$_{28}$Li$_5$[H$_7$P$_8$W$_{48}$O$_{184}$] in water, even at higher LiCl concentrations (see supporting information).

Besides the peaks attributed above to AOT SAM, the PM-RAIRS spectra now displayed the fingerprint of the POM at 1131 cm$^{-1}$, assigned to the $\nu$P-O vibrations (Figure 2).[92] The vibration at 932

cm$^{-1}$ could tentatively by assigned to the terminal WO bonds. However it lies at the border of the optical window of the experimental set-up of the PM-RAIRS instrument (KBr and ZnSe optics and MCT detector).

The mean thickness of the layer was assessed by ellipsometry (see Experimental Section) with several measurements on each substrate to check for the homogeneity: the thickness was found to increase from 1.2 nm for the AOT SAM to 2 nm after the deposition of the POMs. This 0.8 nm increase was obtained in a very reproducible way, whatever the time of immersion. Moreover, the AFM image of the POM layer showed a homogeneous film (figure 1-c). This proves that the POMs were uniformly deposited. Theoretically, a 2.0 nm increase is expected if the POMs have a vertical orientation and a sub-1 nm increase is expected in the case of a horizontal orientation (see scheme 1), because of the porous aspect of a compact monolayer in this case. We thus assume that the $[H_7P_8W_{48}O_{184}]^{33-}$ adopt a horizontal configuration during the electrostatic deposition and rapidly saturate the surface, thus stopping further growing of the thickness. Note that if the rinsing of the AOT SAM was not done properly, the POMs clinged to small aggregates and an inhomogeneous layer was obtained (figure S1 in supporting information), which demonstrates how the PBS rinsing step is crucial. The quality of the seeding layer is thus essential to get a uniformly structured POM layer.

Once optimized with common Au/Si and Au/glass substrates, the procedure was generalized to the deposition of $[H_7P_8W_{48}O_{184}]^{33-}$ onto $^{TS}$Au and Au nanodots to perform electrical measurements. As $^{TS}$Au was very flat (the roughness is around 0.4 nm and there is no grain boundaries), the thickness of the AOT SAM determined by ellipsometry was quite lower (0.7 nm) and the POM layer was measured slightly thinner (0.6 nm). This decrease of the AOT monolayer thickness observed on the $^{TS}$Au could be related to the lower roughness of this SAM (0.60 nm, figure 1-d) compared to the one on Au/glass substrates (roughness around 1.27 nm, figure 1-b). We also note that in the case of SAMs on $^{TS}$Au substrates the PBS rinsing step was not essential, probably because of the higher quality/lower roughness of the bare gold surface (figures 1-d and e). The SAMs on $^{TS}$Au substrates were characterized by XPS, see Experimental Section (Figure 3). The presence of the POMs was inferred from the W4f doublet at 35.7 and 37.9 eV corresponding to the 4f$_{7/2}$ and 4f$_{5/2}$ levels respectively for W(VI) atoms. The P2p peak was also observed at 133.9 eV, in agreement with fully oxidized phosphorous atoms.[93] A S2p doublet was observed at 162.1 ± 0.1 and 163.3 ± 0.1 eV, for the S2p$_{3/2}$ and the S2p$_{1/2}$ signals respectively, corresponding to S bound to Au (75.5 %). An additional contribution was observed at 163.7 ± 0.1 (S2p$_{3/2}$) and 164.8 ± 0.1 eV (S2p$_{1/2}$) attributed to non Au-bonded thiol molecules (24.5%). Finally the protonation of the amino-groups which had driven the electrostatic POM deposition was assessed from the N1s peak, which was decomposed in two almost equal contributions at 400.1 ± 0.1 and 402.1 ± 0.1 eV for the unprotonated (37.7%) and protonated (62.3 %) amino-groups respectively (figure 3).

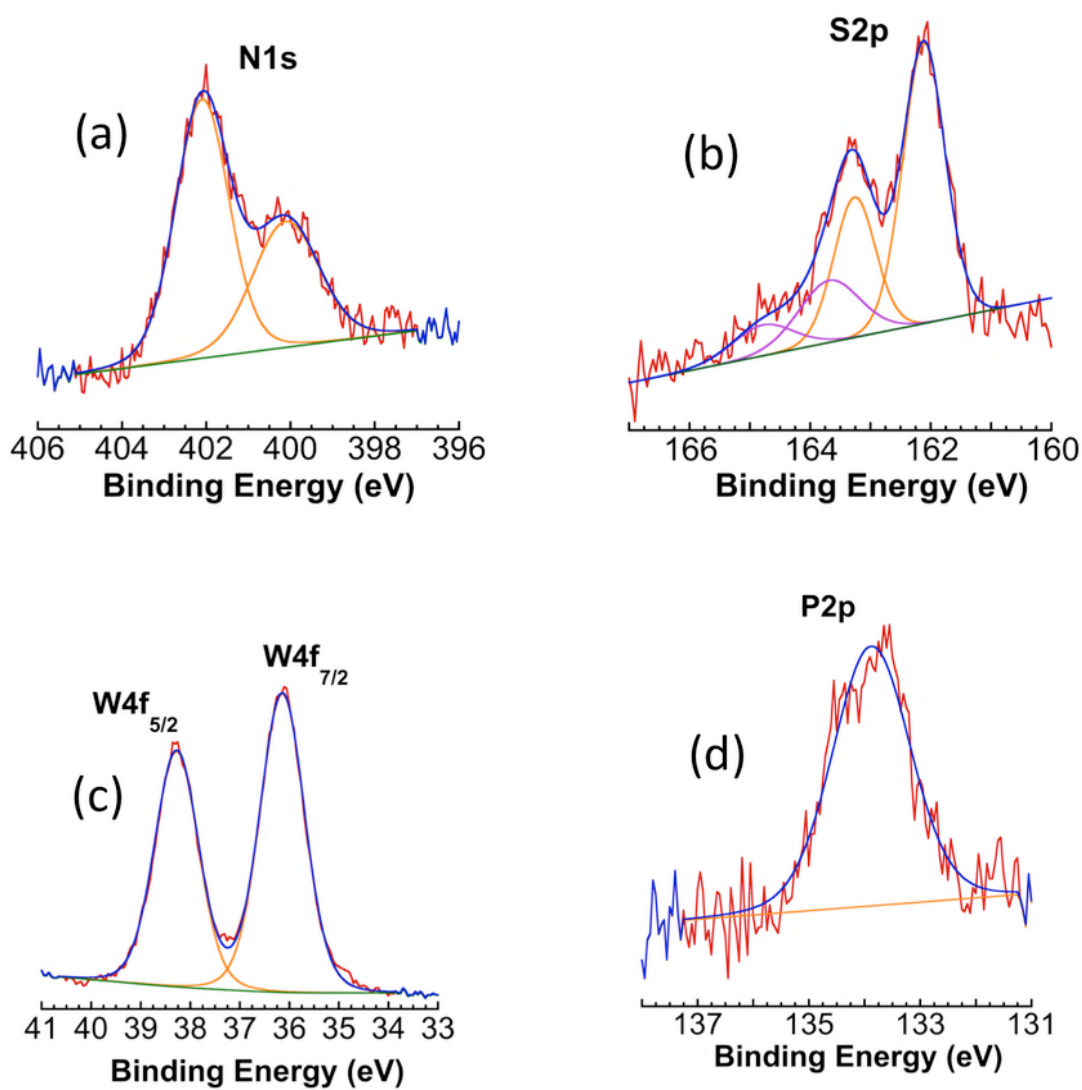

*Figure 3*. XPS characterization of the $[H_7P_8W_{48}O_{184}]^{33-}$ layer onto AOT SAM prepared onto $^{TS}Au$: **(a)** N1s, **(b)** S2p, **(c)** W4f and **(d)** P2p peaks.

## 3. Electrical characterizations.

Electronic transport properties through molecular junctions are impacted by numerous parameters such as contact geometry, molecule orientation, molecular organization, and number of molecules in large junction, for example. Conductance histograms thus become essential to describe electron transport properties in single molecule junctions[94] as well as larger molecular junctions.[95] One way to obtain statistics is to repeat measurements on various single molecule junctions or on various points on the same larger molecular junction. In an alternative way, some of us have recently reported on the use of an array of sub-10 nm Au nanodots to record the conductance of up to a million of alkyl-thiol junctions in a single C-AFM image.[86-87] Analysis of the collected data is used to determine the electronic structure of the molecular junctions.

**POMs on Au-TS electrodes.**

Electronic transport properties of the $[H_7P_8W_{48}O_{184}]^{33-}$ electrostatically deposited onto the AOT SAMs were investigated at the nanoscale by C-AFM. The SAMs grafted on $^{TS}$Au substrate were measured under a nitrogen flow using a platinum tip and at a loading force of 6 nN (see Experimental Section). This conducting tip was placed at a stationary point contact to form $^{TS}$Au-S-C8-$NH_3^+$//$[H_7P_8W_{48}O_{184}]^{33-}$//Pt nanojunctions (here // denotes a non-covalent interface and − a covalent interface). A square grid of 10 × 10 μm is defined with a lateral step of 100 nm. At each point, one I−V curve is acquired (back and forth), leading to the measurements of 100 I−V traces on a given location on the sample. This procedure was repeated at 3 different locations on the sample. The bias was applied on the Au-TS substrate, and the tip was grounded through the input of the current amplifier (figure 4-a). The voltage sweeps (back and forth) were applied from -1.5 to 1.5 V and then from 1.5 to −1.5 V.

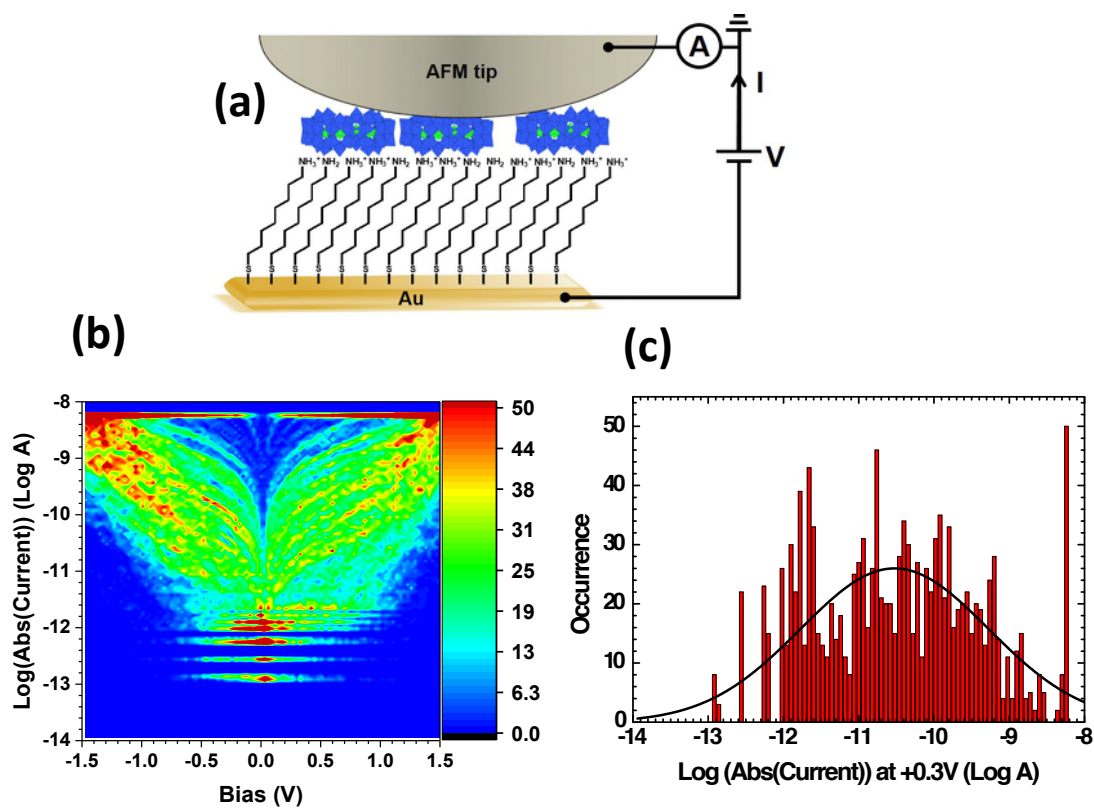

*Figure 4. (a) Scheme of the electronic transport characterization by C-AFM of the $[H_7P_8W_{48}O_{184}]^{33-}$ electrostatically deposited onto AOT SAM. (b) 2D current histogram of 300 I-V curves measured at several locations on the SAM. (c) Current histogram at + 0.3 V adjusted by a log-normal distribution. The higher bar at log I = -8.4 is due to the saturation of the amperemeter and corresponds to the number of counts of current higher than $4\times10^{-9}$ A (i.e. the hidden higher tail of the distribution).*

Figure 4-b shows the 2D histogram of the 300 I-V traces measured on the SAM nanojunctions. The I-V curves are quite symmetric with respect to the voltage polarity. The current histograms at a given bias (for example at +0.3 V in figure 4-c) were fitted by a log-normal distribution, leading to a mean value of log I = -10.5 (i.e. a mean current of 3.2 x $10^{-11}$ A) with a standard variations log-$\sigma \simeq$ 2.5.

**POMs on Gold Nanodots (nanodot molecule junction : NMJ).**
Molecular junctions of the $[H_7P_8W_{48}O_{184}]^{33-}$ electrostatically deposited onto AOT SAM were also fabricated on a large array of single crystal Au nanodot electrodes.[86-87] An array of gold nanodot (10 nm in diameter) electrodes was fabricated by e-beam lithography and lift-off technique.[86-87] Each nanodot is embedded in a highly doped Si substrate (to ensure a back ohmic contact), covered by a SAM of molecules of interest and contacted by the C-AFM tip. Each nanodot/molecules/tip junction contains about 25 POMs (considering a nanodot diameter of 10 nm, a POM diameter of 2 nm, and assuming a close-packing). Figure 5-a shows a typical C-AFM image taken at +0.3 V for 197 Au nanodot-AOT//$[H_7P_8W_{48}O_{184}]^{33-}$//C-AFM tip molecular junctions. The current histogram corresponding to the current image is shown in figure 5-b for voltage + 0.3 V. This histogram was well fitted by a log-normal distributions giving the average current of log I = - 7.29 (i.e. a mean current of 5.1 x $10^{-8}$ A) and a standard deviation log $\sigma \simeq$ 1.5. It may be possible to consider two peaks in the histogram shown in figure 5-b, as observed in previous works on NMJs.[87] However, this feature was not systematically observed for the present samples (see other histograms in the supporting information, figure S3 ). We have mainly observed a tail at low current in the histograms deviating from the log-normal distribution, a feature previously attributed to intermolecular interactions between adjacent molecules in the NMJs.[96-97] This confirms a close packing of the $[H_7P_8W_{48}O_{184}]^{33-}$ molecules in the NMJs. We note that, albeit using the same loading force (6 nN), the currents are higher in the NMJs compared to the SAM on $^{TS}$Au substrates. This has been rationalized in a previous study combining mechanical and electrical characterizations of the molecule/nanodot structure: it was found that the applied pressure is increased in the NMJs due to the smaller contact area (compared to C-AFM on SAM on large metal surface), leading to higher currents.[98] By successive acquisition of the current images at various voltages, a reconstructed I-V curve is obtained for the molecular junction (figure 5-c). The current histograms at the different voltages are given in the supporting information (figure S3).

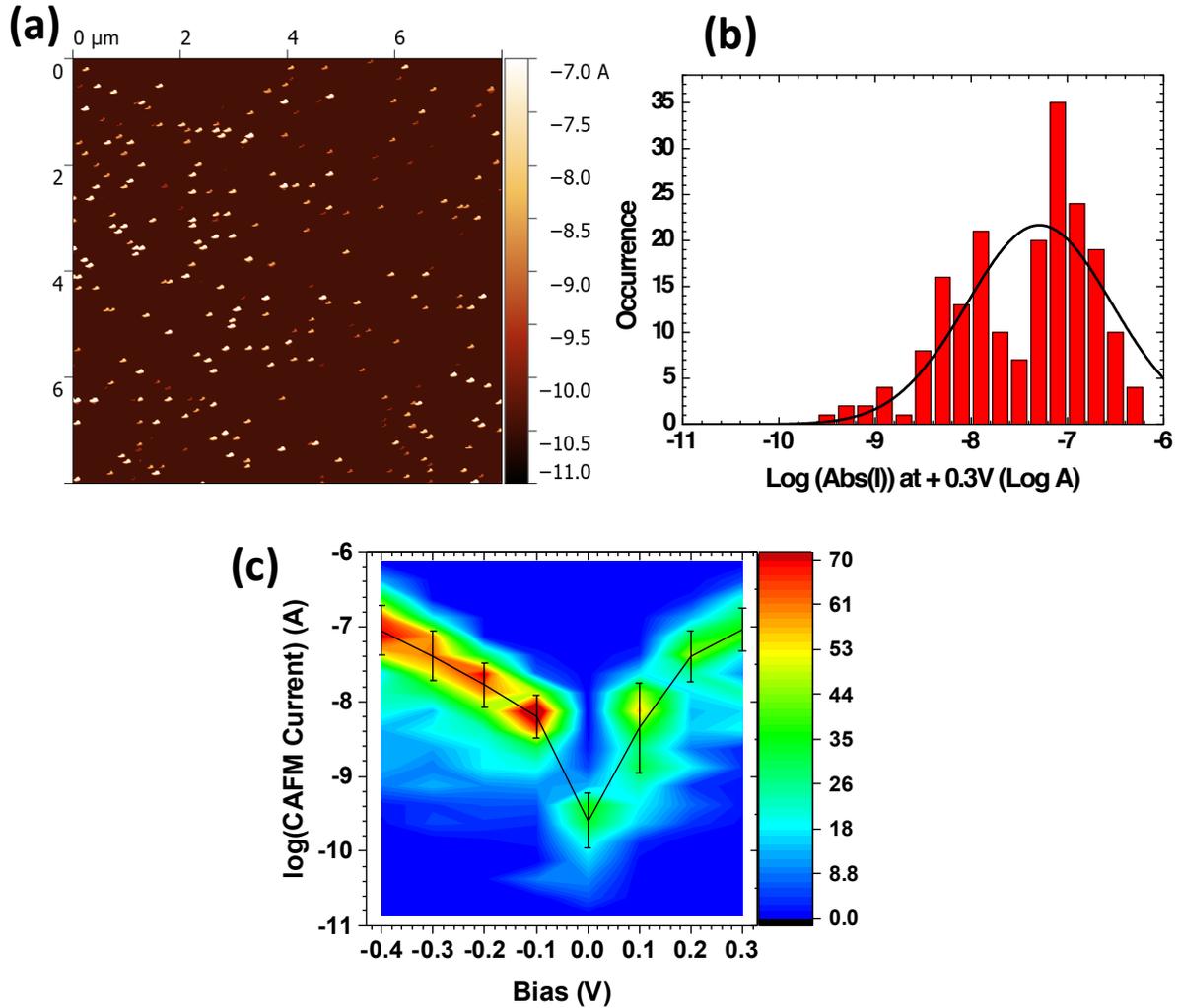

*Figure 5.* *(a)* Current image of 8 x 8 μm at + 0.3 V for a loading force fixed at 6 nN (white spot = positive current) of Au nanodots capped with $[H_7P_8W_{48}O_{184}]^{33-}$ electrostatically deposited onto AOT SAM. *(b)* Corresponding histograms of the current with 197 junctions fitted with a log-normal distribution : average current of log I = -7.29 (i.e. a mean current of 5.1 x $10^{-8}$ A) and a standard deviation log- σ ≃1.5. *(c)* Reconstructed I-V curve from the current image at different voltages: the current value and the incertitude in the curve correspond to the mean current value and standard deviation respectively deduced from the log-normal distribution fitting. The reconstructed I-V curve (black line) is superimposed on the 2D histogram of the current measured at the different voltages on the NMJs. Only a fraction (around 3 %) of the junctions are electrically active (bright spots in figure 5-a) compared to the total number of nanodots on the scanned area (8 x 8 μm, with a nanodot every 100 nm, Experimental Section and images in the supporting information, figure S2). This feature may have several origins (see the supporting information).

# 4. Discussion: molecular junction energetics.

The I-V curves were analyzed with three methods to determine the electronic structure of the molecular junctions, i.e. the energy position of the molecular orbital involved in the electron transport process. The first one consists to fit the I-V curves with the Simmons tunnel equation[99] (see supporting information), the second one is the TVS (transition voltage spectroscopy) method,[100-104] and the third consists to fit the I-Vs with a molecular single energy level model based on the Landauer equation and considering that the electron transfer is dominated by one molecular orbital in the junction ("molecular model", see SI).[105-107] In the first case, the fit gives the effective tunnel energy barrier, $\Phi_T$, seen by electrons to tunnel from one electrode to the other. In the TVS method, a threshold voltage $V_T$ is extracted from the minimum of $(LnI/V^2)$ versus $1/V$ plot. This voltage is related to the energy position of the molecular orbital, $\varepsilon$, relative to the electrode Fermi energy, by $\varepsilon = 0.86eV_T$ with e the electron charge (see supporting information). This energy level is also determined by fitting the "molecular model" and noted as $\varepsilon_0$ in the following. In the present case, since the POMs are strongly accepting molecules, we infer that the molecular orbital involved in the electron transport though the molecular junction is the POM LUMO. Typical examples of these three methods applied on the data shown in figures 4-b and 5-c for the SAM and NMJs respectively are given in figure S4 (supporting information). Figure 6 gives the histograms of the energy levels determined by the three methods for the POMs junctions on $^{TS}$Au substrates. Table I summarizes the values obtained with these three methods on the SAMs and NMJs.

|  | POM SAMs on $^{TS}$Au (eV) | NMJs (eV) |
|---|---|---|
| $\Phi_T$ (Simmons model) | 0.83 ± 0.16 | 0.26 ± 0.33 |
| $\varepsilon$ (TVS method) | 0.72 ± 0.07 | not applicable |
| $\varepsilon_0$ (molecular model) | 0.93 ± 0.15 | 0.24 ± 0.02 |

***Table I.*** *Summary for the values obtained by the three methods on the POM SAMs grafted on Au-TS and NMJs. These values are obtained from the normal distribution of the fitting values presented in Fig. 6 for SAMs and in the supporting information(figure S7) for NMJs. Due to the small numbers of voltages used for the IV measurements, the TVS method is not applicable for NMJs.*

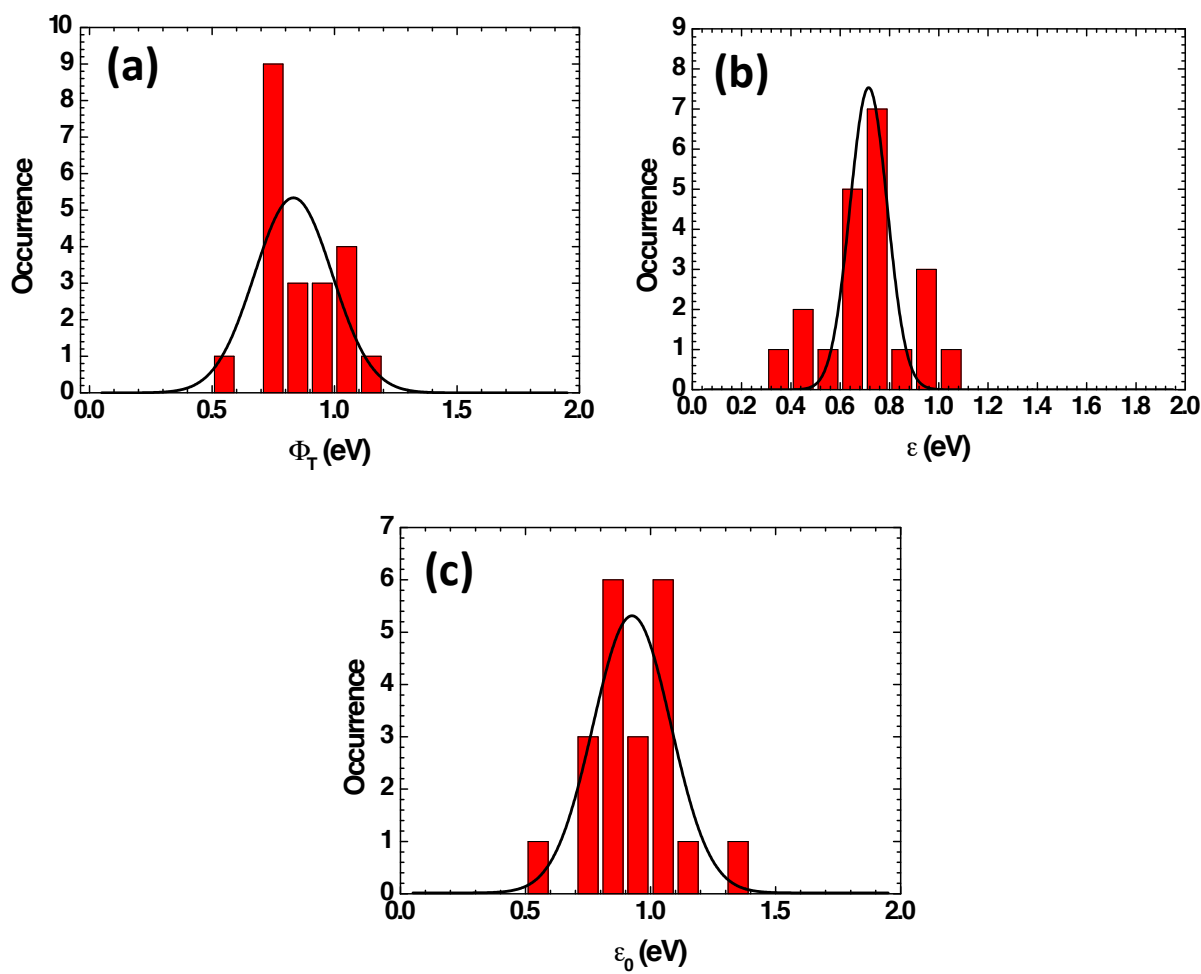

*Figure 6. Statistical distributions of the energy levels deduced by applying the 3 methods on 21 I-V curves arbitrarily taken from data shown in figure 4: **(a)** Simmons model, **(b)** TVS method, **(c)** molecular model. The error bars in Table I are taken form the FWHM of these distributions. Examples of the I-V curve adjustments with the three models and distributions for the other fitting parameters (electron effective mass and electrical surface contact area for the Simmons model) are presented in the supporting information (figures S4, S5 and S6).*

These values are now discussed in term of energetics of the molecular junctions. The first observation is that the $\varepsilon$ values given by the TVS method or by fitting the molecular level are in good agreement as expected since the relation $\varepsilon = 0.86eV_T$ is also based on the single molecular energy model (see SI).[103,108] The measured molecular junctions are constituted of 2 parts: the alkyl spacer and the POM (Scheme 1). The alkyl spacer is electrically insulating with a high HOMO-LUMO gap (7-9 eV),[109-110] while the POM has a smaller gap (4.7 eV from optical absorption). We can consider a staircase diagram for the energetics of the junction as depicted in figure 7 where the LUMO of the alkyl chains (C8) is supposed higher in energy than the LUMO of the POMs. In such a case, the Simmons model considers only a simple rectangular energy tunnel barrier with an effective tunnel barrier $\Phi_T$ between the LUMOs of the alkyl chain and the POM, $\varepsilon_{POM} < \Phi_T < \varepsilon_{C8}$. When a voltage is applied, the Simmons model considers a linear variation of the electric field between the electrodes (figure 7-c). On the contrary, the TVS method used here (with the relationship $\varepsilon = 0.86eV_T$) and the molecular model give the energy of the molecular orbital, considering that the potential drops are mainly located at the molecule (i.e. POM unit)/electrode interface (i.e. through the alkyl spacer and the mechanical contact between the POM and the C-AFM tip) - figure 7-e (for a more detailed discussion of the Simmons model and TVS method, see Refs. [111, 112]). We can consider that the measured $\varepsilon$ and $\varepsilon_0$ (Table I) are reasonable estimate of $\varepsilon_{POM}$. In the WKB (Wentzel-Kramers-Brilloin) approximation, the effective tunnel barrier $\Phi_T$ for such a staircase tunnel barrier (Figure 7) is given by:

$$\sqrt{\Phi_T}(d_{C8}+d_{POM})=\sqrt{\varepsilon_{C8}}d_{C8}+\sqrt{\varepsilon_{POM}}d_{POM} \tag{1}$$

where $d_{C8}$ and $d_{POM}$ are the thicknesses of the C8 alkyl chain and POM, respectively, $d_{C8}$=0.7 nm and $d_{POM}$=0.6 nm (as measured by ellipsometry on the $^{TS}$Au substrates). Since the values for $\varepsilon_{POM}$ (i.e. $\varepsilon$ or $\varepsilon_0$) and $\Phi_T$ are quite similar (Table I), we conclude that the LUMO of the alkyl spacer, $\varepsilon_{C8}$, and the POM, $\varepsilon_{POM}$, are roughly similar (alkyl spacer and POM LUMOs energetically aligned). This feature is consistent with the value $\varepsilon_{C8}$ of ca. 1 eV previously determined for the LUMO of C8 (pure alkyl chain, no amine) SAM on Au by TVS[104] and IPES.[113] We hypothesis that the dipole at the AOT/POM interface (positive charge at the AOT amine end-group, negative charge on the POM side) can induced this alignment by shifting downstairs (upstairs, respectively) the energetics of C8 and POM, respectively.

The second observation is that the values for NMJs are smaller than for SAMs on $^{TS}$Au. Again this is consistent with the higher currents measured on NMJs and with previous report on the mechanical behaviors on NMJs (higher pressure at the same loading force induces a reduction of $\varepsilon$).[98] We previously reported a reduction by a factor about 2 of the TVS value ($\varepsilon$) for a C8 monolayer on NMJs[98] compared to a C8 SAM on golf surface,[104] both measured by C-AFM at the same loading force. The fact that, for NMJs, we still have $\Phi_T \approx \varepsilon_0$ means that the same conclusion (energy level alignment of the alkyl spacer and POM LUMOs) holds for NMJs.

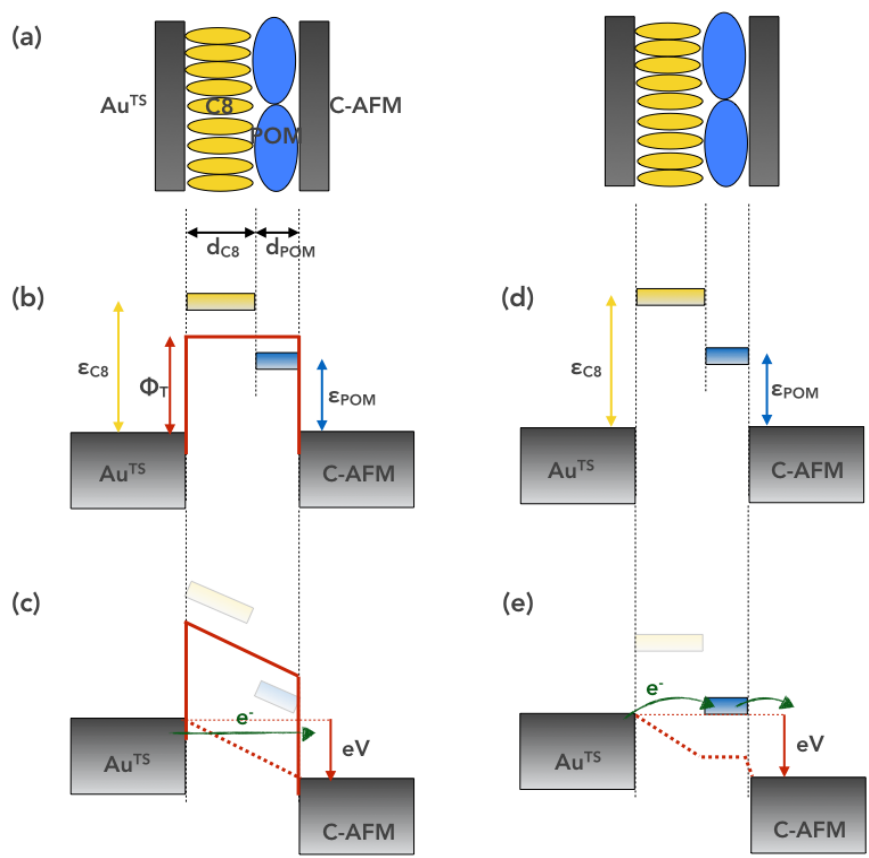

*Figure 7. (a) Scheme of the staircase energetic diagram of the Au-AOT//POM//C-AFM tip junction. (b-c) Effective tunnel barrier $\Phi_T$ determined with the Simmons model at 0 V (b) and at a voltage V (c). (d-e) Determination of the energy position of the POM LUMO, $\varepsilon_{POM}$, by the TVS method and molecular model at 0 V (d) and at a voltage V (e).*

## 5. Conclusion.

In this contribution, we have described a reliable procedure for the electrostatic immobilization of POMs onto positively charged SAMs of amino-alkylthiols onto Au. Uniformly structured monolayers of POMs have been directly obtained onto $^{TS}$Au, while a rinsing methodology at the SAM formation step had to be optimized in the case of usual Au/glass or Au/Si substrates of higher roughness. Controlling the morphology/organization of the POM layers is essential to ensure reproducible properties. This procedure is easy to extend to other POMs, especially those with high electron-storage abilities. We have thus recorded the first electrical data on molecular junctions involving the large $[H_7P_8W_{48}O_{184}]^{33-}$ POM, analyzed these data by various methods and experimentally determined the energy position of the lowest unoccupied molecular orbital of the POM (with respect of the metal electrode Fermi energy), which governs electron transport through these molecular devices.

# Experimental section.

**Materials**

*Chemicals and solvents* were purchased from Sigma-Alrich and used as received. $K_{28}Li_5[H_7P_8W_{48}O_{184}]$ was prepared as described in the literature[92] and its purity checked by $^{31}P$ NMR and electrochemistry (figures S8-S10).

The *Au/Si substrates*, purchased from Sigma-Aldrich, consisted of a 100 nm gold layer adhered to the silicon wafer by means of a titanium adhesion layer. The substrates were treated 2 min in pure sulfuric acid and rinsed abundantly with water and isopropanol and dried under $N_2$. Before immersion in the thiol solution, the substrates were exposed to UV-ozone during 20 minutes and rinsed with ethanol.

*Au/glass substrates*, coated successively with a 50 Å thick layer of chromium and a 200 nm thick layer of gold, were purchased from Arrandee (Werther, Germany). To ensure a good crystallinity of the gold top layer, the substrates were annealed in a butane flame, afterwards exposed to UV-ozone during 20 minutes and rinsed with ethanol prior to immersion in the thiol solution.

*Template-stripped Au substrates.* Very flat $^{TS}$Au surfaces were prepared according to the method reported by the Whiteside group.[85] In brief, a 300−500 nm thick Au film is evaporated on a very flat silicon wafer covered by its native $SiO_2$ (RMS roughness of 0.4 nm), which was previously carefully cleaned by piranha solution (30 min in 2:1 $H_2SO_4/H_2O_2$ (v/v); Caution: Piranha solution is exothermic and strongly reacts with organics), rinsed with deionized (DI) water, and dried under a stream of nitrogen. A clean glass piece (ultrasonicated in acetone for 5 min, ultrasonicated in 2-propanol for 5 min, and UVirradiated in ozone for 10 min) is glued (UV polymerizable glue) on the evaporated Au film and mechanically stripped with the Au film attached on the glass piece (Au film is cut with a razor blade around the glass piece). This very flat (RMS roughness of 0.4 nm, the same as the $SiO_2$ surface used as template) and clean template-stripped AuTS surface is immediately used for the formation of the SAM.

*Gold nanodots fabrication.* The fabrication and detailed characterization of these nanodot arrays have been reported elsewhere.[86] For e-beam lithography, an EBPG 5000 Plus from Vistec Lithography was used. The (100) Si substrate (resistivity = $10^{-3}$ Ω.cm) was cleaned with UV-ozone and native oxide etched before resist deposition. The e-beam lithography has been optimized by using a 45 nm thick diluted (3:5 with anisole) PMMA (950 K). For the writing, an acceleration voltage of 100 keV was used, which reduces proximity effects around the dots, compared to lower voltages. The beam current to expose the nanodots was 1 nA. The conventional resist development / e-beam Au evaporation (8 nm) / lift-off processes were used. Immediately before evaporation, native oxide is removed with dilute HF solution to allow good electrical contact with the substrate. At the end of the process, these nanodots were covered with a thin layer of $SiO_2$ that was removed by HF at 1% for 1 mn prior to SAM deposition. Spacing between Au nanodots was set to 100 nm (see images in the supporting information, figure S2).

**Characterization techniques.**

*Ellipsometry* Ellipsometry measurements were performed on the Au/Si substrates with a monowavelength ellipsometer SENTECH SE 400 equipped with a He-Ne laser at λ = 632.8 nm. The incident angle was 70°. As the optical indices of the bare gold substrate change from one sample to another, the ns and ks values of the sample with bare Au were systematically measured just before immersion in thiol solution. The ns and ks values were around 0.2 and 3.5 respectively. For the 8-amino-1-octanethiol SAM, typical optical indices of an organic monolayer were used ($n_s$ = 1.5, $k_s$ = 0). Finally, $n_s$ = 1.48 and $k_s$ = 0 were used for the layer of POMs.[79] At least 6 measurements were performed on a same sample in different zones, to check the homogeneity of the layer. A mean value for the thickness was calculated when the standard deviation was equal or lower than 0.2 nm.

*X-ray photoelectron spectroscopy* XPS analyses were performed on using an Omicron Argus X-ray photoelectron spectrometer. The monochromated $AlK_α$ radiation source ($hν$ = 1486.6 eV) had a 300 W electron beam power. The emission of photoelectrons from the sample was analyzed at a takeoff angle of 90° under ultra-high vacuum conditions ($\leq 10^{-10}$ Torr). Spectra were carried out with a 100 eV pass energy for the survey scan and 20 eV pass energy for the P2p, W4f, C1s, O1s, N1s, S2p regions. Binding energies were calibrated against the $Au4f_{7/2}$ binding energy at 84.0 eV and element peak intensities were corrected by Scofield factors.[114] The peak areas were determined after subtraction of a linear background. The spectra were fitted using Casa XPS v.2.3.15 software (Casa Software Ltd., U.K.) and applying a Gaussian/Lorentzian ratio G/L equal to 70/30.

*Polarized modulated reflexion absorption infrared spectroscopy (PM-RAIRS)*

PM-RAIRS measurements were performed on Au/Si or Au/glass substrates using a Nicolet Nexus 5700 FT-IR spectrometer equipped with a nitrogen-cooled HgCdTe wide band detector. Infrared spectra were recorded at 8 $cm^{-1}$ resolution, by co-addition of 128 scans. A ZnSe grid polarizer and a ZnSe photoelastic modulator were placed prior to the sample in order to modulate the incident beam between p and s polarizations (HINDS Instruments, PM90, modulation frequency = 36 kHz). The sum and difference interferograms were processed and underwent Fourier-transformation to yield the PM-RAIRS signal which is the differential reflectivity ($\Delta R/R°$) = ($R$p-$R$s)/($R$p+$R$s), where Rp is the signal parallel to the incident plane while Rs is the perpendicular contribution. The measurements were done at two different voltages applied to the modulator ZnSe crystal to optimize the sensitivity.

*Atomic force microscopy.*

AFM images were recorded on Au/glass substrates using a commercial AFM (NanoScope VIII MultiMode AFM, Bruker Nano Inc., Nano Surfaces Division, Santa Barbara, CA) equipped with a 150 × 150 × 5 μm scanner (J-scanner). The substrates were fixed on a steel sample puck using a small piece of adhesive tape. Images were recorded in peak force tapping mode in air at room temperature (22−24 °C) using oxide-sharpened microfabricated $Si_3N_4$ cantilevers (Bruker Nano Inc., Nano Surfaces Division, Santa Barbara, CA). The spring constants of the cantilevers were measured using the thermal noise method, yielding values ranging from 0. 4 to 0.5 N/m. The curvature radius of silicon nitride tips was about 10 nm (manufacturer specifications). The raw data were processed using the imaging processing software NanoScope Analysis, mainly to correct the background slope between the tip and the surfaces.

*C-AFM measurements.* Conducting atomic force microscopy (C-AFM) was performed under a flux of $N_2$ gas (ICON, Bruker), using a solid metal probe in platinum (tip radius of curvature less than 20 nm, force constant 0.3 N/m, reference: RMN-12Pt400B from Bruker). The tip loading force on the surface was fixed at 6 nN thanks to force−distance curves with the controlling software of the ICON. The choice of the loading force of 6 nN is a tradeoff between too low or unstable currents and too high deformation or damage of the organic monolayer.[115] Previous works both on SAMs on large electrodes[116] and NMJs[98] have estimated a deformation below 0.1 nm at 6 nN for C8 alkyl chains (as used here). These values remain smaller than the accuracy of the ellipsometry measurements (see above) used to determine the monolayer thickness, which is further used as a parameter in our model to analyze the current-voltage curves. Images were acquired with a sweep frequency of 0.5 Hz and the voltage was applied on the substrate.

**Preparation of the seeding 8-amino-1-octanethiol SAM**

The gold substrate was immersed in a solution of 8-amino-1-octanethiolhydrochloride in absolute ethanol AE (1mg in 50 mL, $10^{-4}$ mol.L$^{-1}$) during 24h, protected from light. Then the surface was rinsed with AE, plunged into an AE bath during 5 min, sonicated 5 min in a new AE bath, rinsed with AE and dried with a $N_2$ flow. The dried substrate was treated by immersion in a PBS solution during 3h (0.01M, pH=7.4), then submitted to several successive rinsing steps: (i) distilled water flow, (ii) sonication 5 min in distilled water, (iii) distilled water then AE flow, before being dried under $N_2$.

**Immobilization of $K_{28}Li_5[H_7P_8W_{48}O_{184}]$**

A 2 mol.L$^{-1}$ LiCl solution in water was prepared (0.8 g in 10 mL) then 136.1 mg of $K_{28}Li_5[H_7P_8W_{48}O_{184}]$ were added ($10^{-3}$ mol.L$^{-1}$). The thiol modified gold substrate was immersed in the clear solution during 1 hour. Finally, the substrate was rinsed with a distilled water flow, AE flow and dried under $N_2$. Decreasing (20 min) or increasing (24 h) the incubation time was shown (see supporting information) to have little effect on the layer thickness according to ellipsometry measurements. The films are at least stable in the course of the electrical and physical characterizations (i.e. days to weeks). All the measurements have been carried out at the solid state and at variance with electrochemical studies carried out in solution, no leaching of the POMs is feared. We also questioned the stability of such electrostatically assembled POM layer in solution. The modified gold substrate (Au/glass) was placed in a 1 M solution of $(NBu_4)_4NPF_6$ in acetonitrile and sonicated for a few hours, then rinsed with acetonitrile. Despite these quite harsh conditions, ellipsometry did not reveal any significant modification of the thickness, which led us to conclude that POMs were not significantly released from the substrate. This is tentatively ascribed to the high total charge of the POMs.

**Supporting information.** Additional AFM images, current histograms for NMJs, descriptions and equations of the models, examples of fitted I-V curves, histograms of fitted Simmons model parameters (effective mass, electrical contact area), cyclic voltamogram and NMR spectrum of $K_{28}Li_5[H_7P_8W_{48}O_{184}]$..

**Conflicts of interest.** There are no conflicts to declare.

**Acknowledgements.** K.D.F. thanks the program PER-SU of Sorbonne Universités for funding. We thank the French RENATECH network (IEMN clean-room). The authors would like to acknowledge IMPC (Institut des Matériaux de Paris Centre, FR CNRS 2482) at Sorbonne Université and the C'Nano projects of Région Ile-de-France for Omicron-XPS apparatus funding.

# Charge transport through redox active [H$_7$P$_8$W$_{48}$O$_{184}$]$^{33-}$ polyoxometalate self-assembled onto gold surfaces and gold nanodots.

K. Dalla Francesca,[1] S. Lenfant,[2] M. Laurans,[1] F. Volatron,[1] G. Izzet,[1] V. Humblot,[3] C. Methivier,[3] D. Guerin,[2] A. Proust,[1,*] D. Vuillaume[2,*]

*1) Sorbonne Université, CNRS, Institut Parisien de Chimie Moléculaire, IPCM, 4 Place Jussieu, F-75005 Paris, France.*
*2) Institute for Electronics Microelectronics and Nanotechnology (IEMN), CNRS, Lille Univ., Av. Poincaré, F-59652 Villeneuve d'Ascq, France.*
*3) Sorbonne Université, CNRS, Laboratoire de réactivité de surface, LRS, 4 Place Jussieu, F-75005 Paris, France.*
* Corresponding authors: anna.proust@upmc.fr, dominique.vuillaume@iemn.fr

## SUPPORTING INFORMATION

### 1. Formation of POM aggregates in the absence of PBS rinsing.

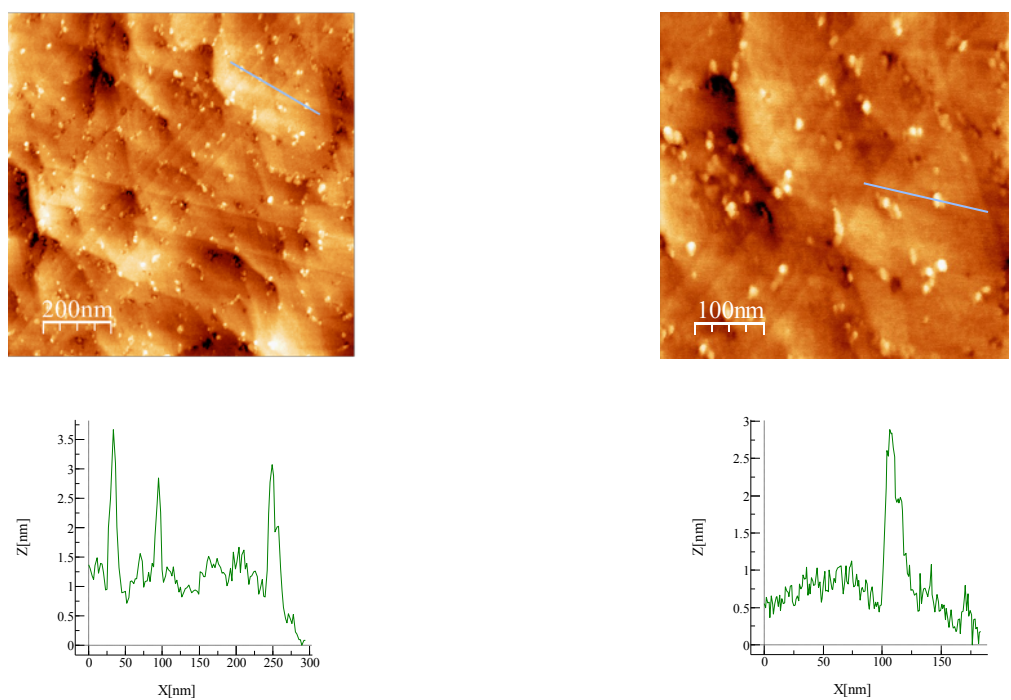

**Figure S1.** 2x2 µm and 1x1 µm AFM images and height profiles of a slightly rinsed AOT modified gold substrate (without PBS treatment) immersed in a 1 mM solution of H$_3$[PW$_{12}$O$_{40}$] in water. The POMs tend to cling to the physisorbed thiol molecules to form bigger aggregates (around 2.5 nm according to the line profiles recorded on the aggregates).

## 2. POM incubation

The samples were prepared by incubation of the AOT functionalized gold substrate in a 1 mM POM solution for one hour. We have varied the incubation time from 20 mn to 1h and 24h without observing any significant modification of the layer thickness as follows by ellipsometry. Regarding the POM solution concentration, we have started with 1mM which is the concentration often used. As the 0.8 nm increase of the thickness was slightly lower than expected (POM thickness of about 1.0 nm) we did not try to lower the POM solution concentration but rather we tried to increase it. Unfortunately, we were limited by the low solubility of the POM and longer immersion times (until 24 h) led to exactly the same layer features and significantly more concentrated solutions were not possible to obtain because of the low solubility of the $K_{28}Li_5[H_7P_8W_{48}O_{184}]$ in water, even at higher LiCl concentrations.

## 3. Nanodots

Figure S2 shows the scanning electron microscope images of the nanodot array after the fabrication. We clearly observed a large and regular array of 10 nm in diameter nanodots separated by 100 nm. The fact that only a fraction (around 3 %) of the junctions are electrically active (bright spots in figure 5-a in the main text) may have several origins: nanodots removed during the chemical treatments or removed during the C-AFM measurements (if not well embedded in the Si substrate). We can discard the removal during the C-AFM measurements since we have not observed a significant variation of the number of active NMJs during successive scanning of the samples to measure the current at several voltages as shown in the current histograms (see below figure S3). It is likely that the nanodots were not well embedded onto the substrate for the present case and were removed during the grafting process in solution.

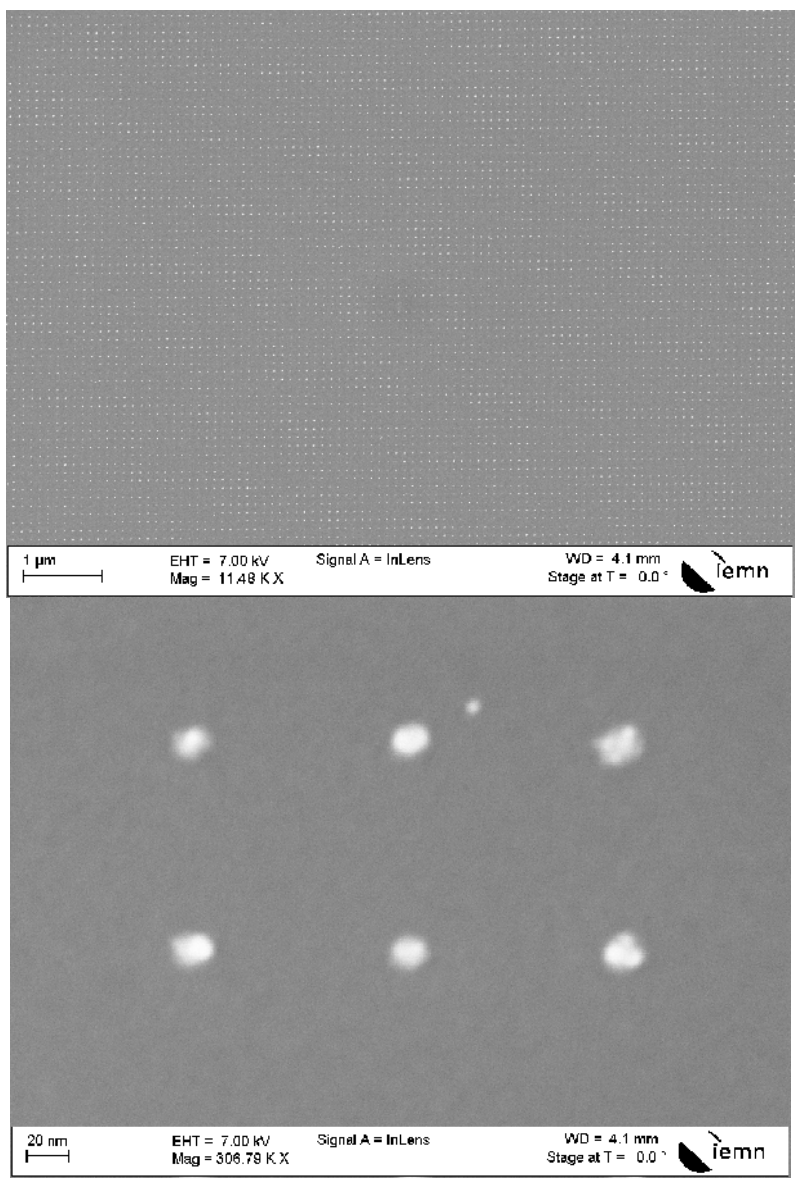

**Figure S2.** Scanning electron microscope images of the nanodots arrays at two magnifications.

## 3. Current histograms at different voltages measured on NMJs.

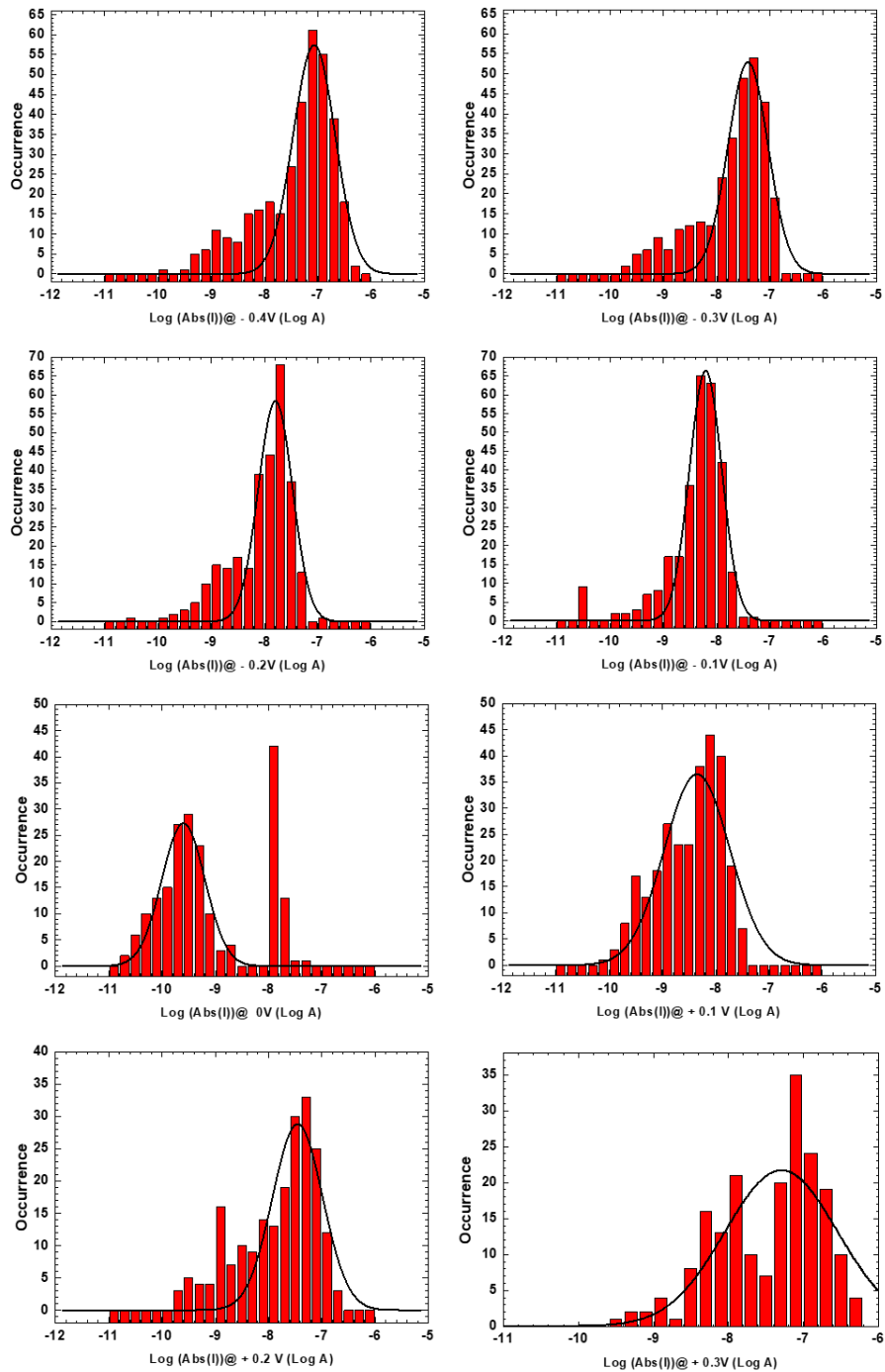

**Figure S3**. Histograms of the currents measured on NMJs at several voltages (as indicated in the x-axis legend). The black lines are fit with a log-normal distribution. Mean values and standard deviations are given in the table S1.

| Applied Voltage (V) | < log I > (log A) | log σ |
|---|---|---|
| -0.4 | -7.05 | 0.66 |
| -0.3 | -7.41 | 0.74 |
| -0.2 | -7.78 | 0.64 |
| -0.1 | -8.20 | 0.6 |
| 0 | -9.59 | 0.72 |
| 0.1 | -8.35 | 1.2 |
| 0.2 | -7.46 | 0.90 |
| 0.3 | -7.29 | 1.50 |

**Table S1.** Mean values and standard deviations of the log-normal distributions fitted on the histograms of the measured currents shown in Fig. S2.

## 5. I-V curves adjustment with the Simmons model.

The expression of the tunnel current through a potential barrier is given by Simmons:[1]

$$I(V) = S \frac{e}{4\pi h d^2} \left[ (2\Phi_T - eV)\exp\left(-\frac{4\pi d \sqrt{m(2\Phi_T - eV)}}{h}\right) - (2\Phi_T + eV)\exp\left(-\frac{4\pi d \sqrt{m(2\Phi_T + eV)}}{h}\right) \right] \quad (S1)$$

with e the electron charge, h the Planck's constant, d the thickness of the tunneling barrier, $\Phi_T$ the energy barrier height, V the voltage applied across the junction, m the effective mass of the electron, I the current and S the electrical contact surface area (smaller than the geometrical contact area, due to defects in the SAM and on the substrate, e.g. roughness, grain boundaries,…).[2-3] The electron effective mass m is $m = m_r m_0$ with $m_0$ the mass of the electron and $m_r$ the reduced mass.

Adjustments of the measured I-V curves are systematically done for the positive and negative bias. The parameter d corresponding to the total thickness of the monolayer determined by ellipsometry (1.3 nm, see main text) is fixed and the three other parameters ($\Phi_T$, $m_r$ and S) are the fitting parameters.

Figure S4 gives 3 examples of the fits for SAMs on ᵀˢAu surfaces. The histograms for $\Phi_T$, $m_r$ and S are given in Figs. 6 (main text), S5 and S6, respectively. The average S value corresponds to a C-AFM tip electrical contact with a diameter of about 0.45 nm. Albeit the real value of the C-AFM tip contact area is difficult to estimate (it depends on the loading force, Young modulus of the SAM)[4] this value seems reasonable considering that the tip has a radius of 20 nm and the low loading force (6nN).

## 6. Transition voltage spectroscopy (TVS).

The I–V curves are analyzed by the TVS method.[5-7] In brief, the I– V data are plotted as $Ln(I/V^2)$ vs. $1/V$. A minimum in this curve corresponds to a transition from a direct tunneling electron transport through the molecules to a resonant tunneling via a frontier molecular orbital (LUMO or HOMO). The energy position ε of the orbital involved in the transport mechanism with respect to the Fermi energy of the metal electrode is given by :

$$\varepsilon = 2 \frac{e|V_{T+}V_{T-}|}{\sqrt{V_{T+}^2 + 10|V_{T+}V_{T-}|/3 + V_{T-}^2}} \quad (S2)$$

with e the electron charge, $V_{T+}$ and $V_{T-}$ the voltage of the minima of the TVS plot at positive and negative voltages, respectively.[8] Eq. S2 reduces to:

$$\varepsilon = 0.86\, e\, V_T \quad (S3)$$

when the I-V curves are symmetric with respect of the applied voltages (our case here) and $V_{T+}=V_{T-}$.

Figure S4 shows 3 examples of TVS plots (on the same IV curves fitted with the Simmons equation). The corresponding histograms of ε are given in Fig. 6 (main text).

## 7. Molecular single energy level model.

A simple analytical model to describe electron transport through a molecular junction is the single energy level model.[9-13] This model is based on the following assumptions: (i) the transport is phase coherent (tunneling mechanism),

(ii) the current is dominated by a single energy level, $\varepsilon_0$, of the molecule (HOMO or LUMO) in the voltage range considered to fit the experimental I-V curves, and (iii) the voltage drops exclusively over the contacts which are described by the coupling constants $g_1$ and $g_2$.

$$I(V) = N\frac{8e}{h}\frac{g_1 g_2}{g_1+g_2}\left[\arctan\left(\frac{2\varepsilon + eV\frac{g_1-g_2}{g_1+g_2}+eV}{2(g_1+g_2)}\right) - \arctan\left(\frac{2\varepsilon + eV\frac{g_1-g_2}{g_1+g_2}-eV}{2(g_1+g_2)}\right)\right] \quad (S4)$$

with h the Planck's constant and N the number of molecules in the junctions. The fitting parameters are $\varepsilon$, $g_1$ and $g_2$. Since the exact number of molecules in the junction is not known (albeit weak, e.g. <25 in the NMJs, see main text) we fixed N=1, but we have observed that the exact value of N does not have a drastic influence on $\varepsilon$, only on the coupling parameters $g_1$ et $g_2$. Albeit the obtained values seem reasonable (in the range 0.1 - few tens of meV), it is known that these fitting parameters are also very sensitive to the I-V curve "quality" (e.g. noise, sudden jumps in the IV,…)[13] while the value of $\varepsilon$ is not, consequently we have not discussed these values in more detail.

Figure S4 shows 3 examples of these fits on the same IVs (SAMs on $^{TS}$Au) also analyzed with the Simmons equation and the TVS method. The corresponding histogram of $\varepsilon_0$ is shown in Fig. 6 (main text).

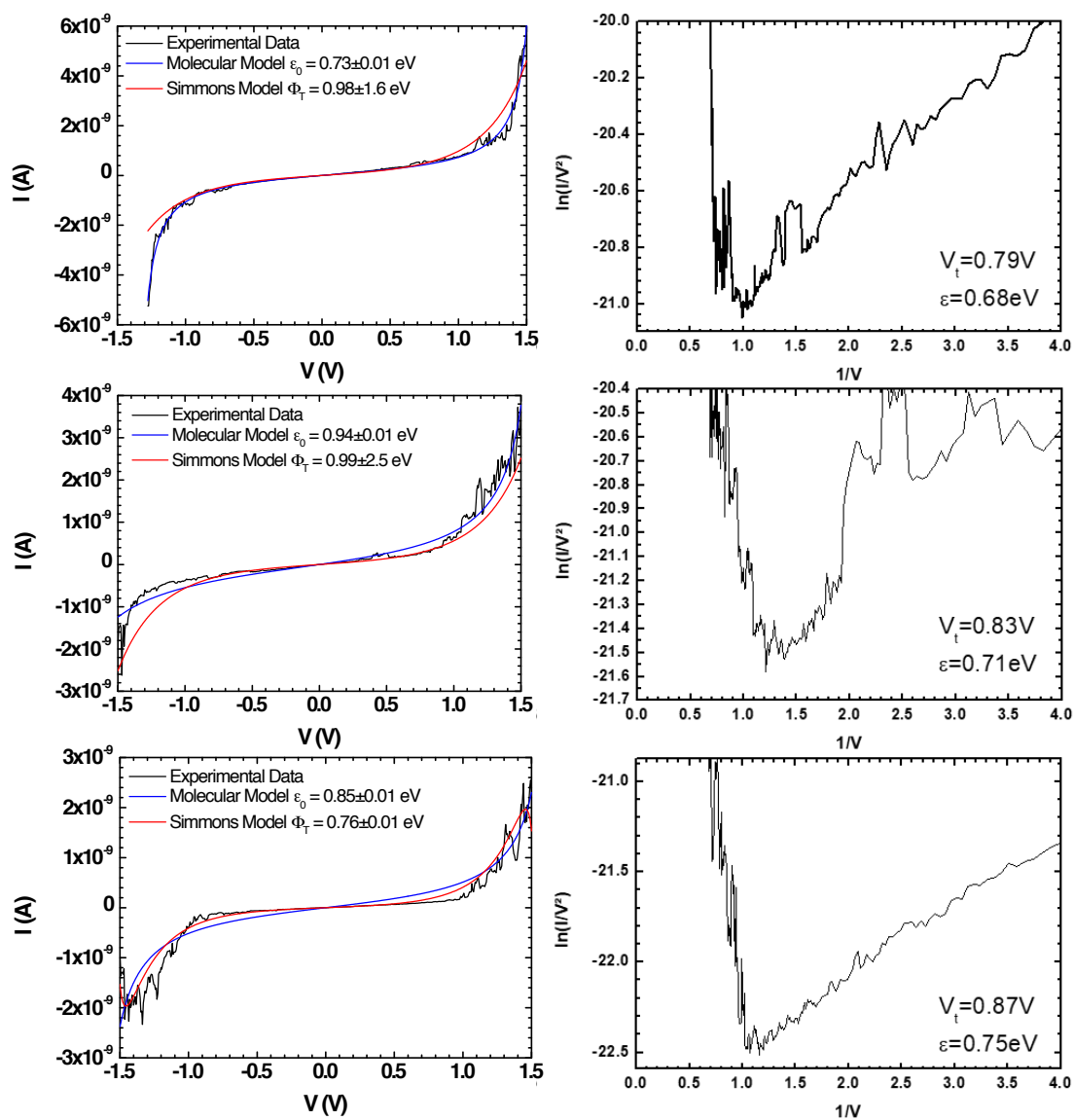

**Figure S4.** Typical examples of the fit (Simmons and molecular models) of 3 different I-V curves measured by C-AFM on the SAM on $^{TS}$Au substrates (left), and corresponding plot of the TVS method (right).

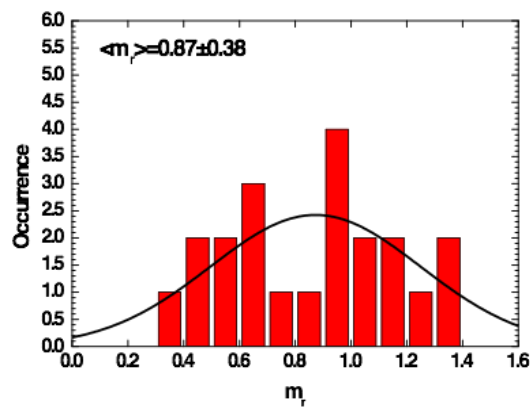

**Figure S5.** Histograms of the effective reduced mass from Simmons model fitted on I-V curves measured for POM SAMs on $^{TS}$Au electrodes. Black curve: Gaussian fit.

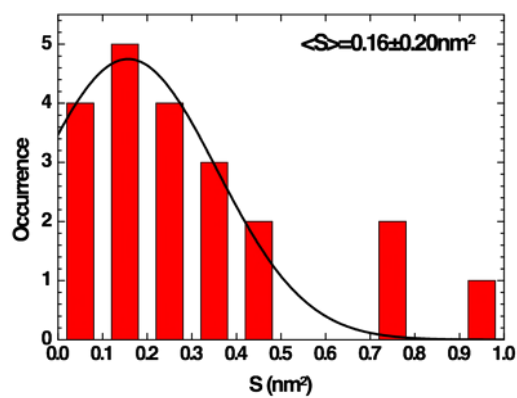

**Figure S6.** Histograms of the electrical contact area from Simmons model fitted on I-V curves measured for POM SAMs on $^{TS}$Au electrodes. Black curve: Gaussian fit.

## 8. Application of the Simmons and molecular models on I-V measured on NMJs.

Figure S7 shows a typical example of the Simmons model and molecular model fitted on the same I-V curve measured on NMJ. For the Simmons model, the thickness was fixed at 1.3 nm as for SAM on $^{TS}$Au since it is not easy to measure the POM thickness on the Au nanodots.[14]

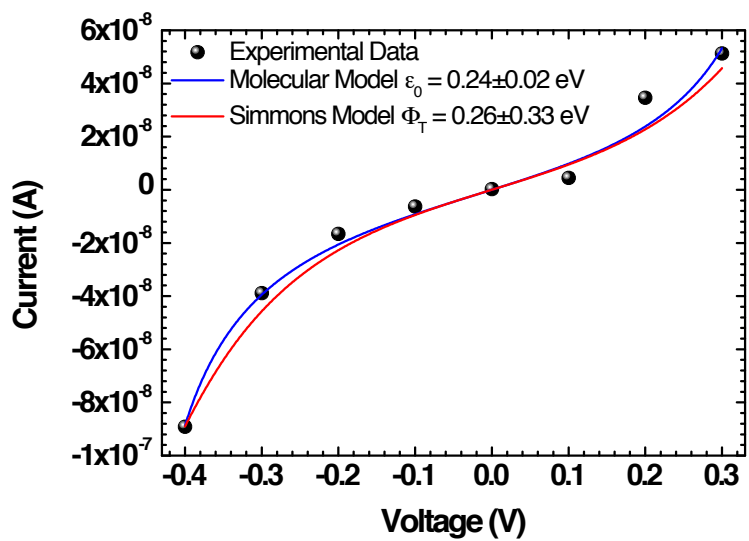

**Figure S7.** I-V curve (same data as in Fig. 5-c, mean current vs. voltages from histograms in Fig. S2) measured on POM NMJ, and fits with Simmons equation and molecular model.

## 9. Characterization of $[H_7P_8W_{48}O_{184}]^{33-}$

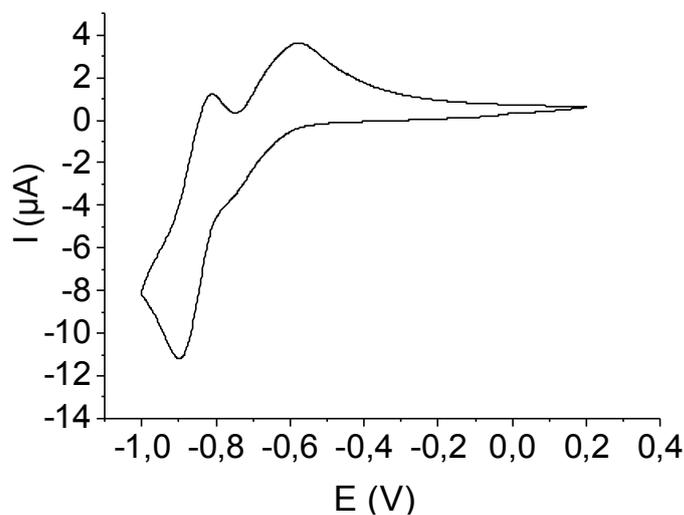

**Figure S8.** Cyclic voltammogram of $K_{28}Li_5[H_7P_8W_{48}O_{184}]$ ($10^{-4}$ M) in $CH_3COOH/CH_3COOLi$ 0.1M in water at a scan rate 10 mV.s$^{-1}$. Working electrode = glassy carbon, reference = calomel saturated electrode (SCE), counter-electrode = platinum wire. It is compliant with those published in by Keita et al.[15]

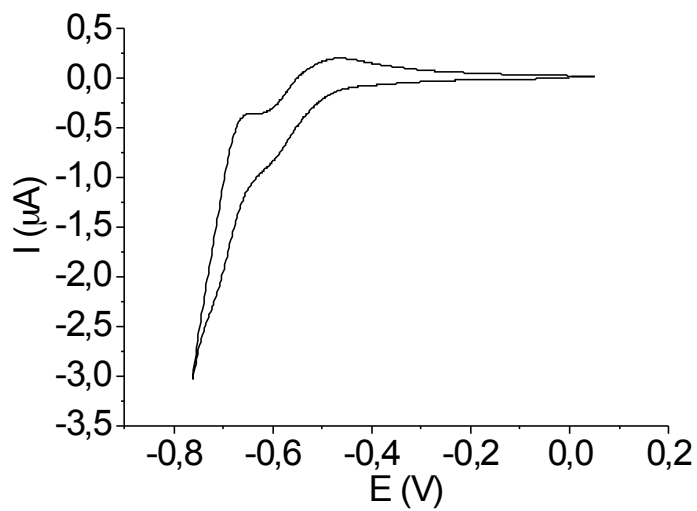

**Figure S9.** Cyclic voltammogram of $K_{28}Li_5[H_7P_8W_{48}O_{184}]$ ($10^{-4}$ M) in $CH_3COOH/CH_3COOLi$ 0.1M in water at a scan rate 10 mV.s$^{-1}$. Working electrode = Au microelectrode, reference = calomel saturated electrode (SCE), counter-electrode = platinum wire.

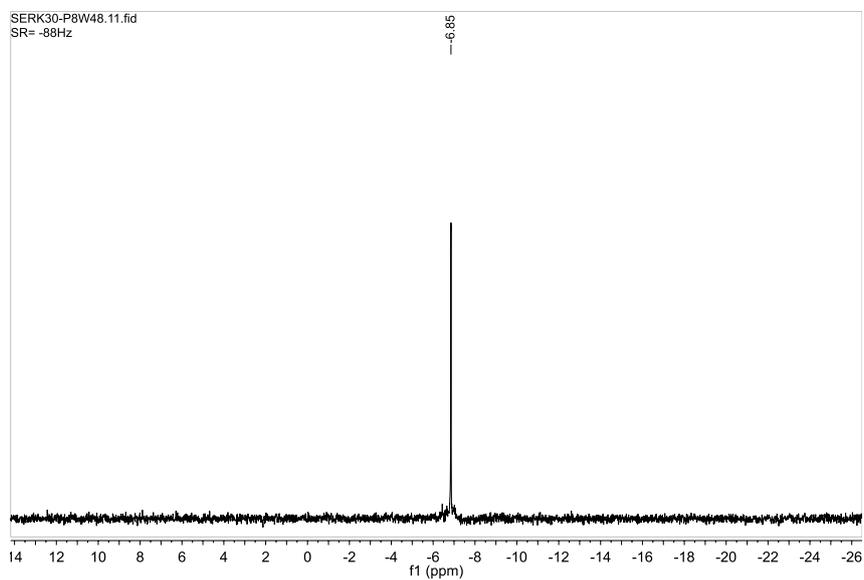

**Figure S10.** $^{31}P$ (121 MHz) NMR spectrum of $K_{28}Li_5[H_7P_8W_{48}O_{184}]$ in LiCl 1M / $D_2O$